\documentclass[11pt]{article}
\usepackage{amsmath,amssymb}
\usepackage{graphicx}
 \allowdisplaybreaks
 \newcommand{\be}{\begin{equation}}
 \newcommand{\ee}{\end{equation}}
 \newcommand{\bea}{\begin{eqnarray}}
 \newcommand{\eea}{\end{eqnarray}}
 
 \newcommand{\nn}{\nonumber}
 \newcommand{\td}{\tilde}
 \newcommand{\wtd}{\widetilde}
 \newcommand{\pd}{\partial}
 \newcommand{\one}{{\bf 1}}

 \newcommand{\cA}{{\cal A}}

 \newcommand{\cO}{{\cal O}}

 \newcommand{\bpsi}{\bar\psi}
 
\long\def\symbolfootnote[#1]#2{\begingroup%
\def\thefootnote{\fnsymbol{footnote}}\footnote[#1]{#2}\endgroup}

\newcommand{\aei}{\it Max Planck Institute for Gravitational Physics
(Albert Einstein Institute)\\ Am M\"uhlenberg 1, 14476 Golm,
Germany}

\newcommand{\auth}{Jianwei Mei}

\begin{document}


\begin{center}

~\vspace{20pt}

\centerline{\Large\bf Spinor Fields and Symmetries of the
Spacetime}

\vspace{25pt}


\auth



\vspace{10pt}{\aei}

\vspace{2cm}

\underline{ABSTRACT}

\end{center}

In the background of a stationary black hole, the ``conserved
current" of a particular spinor field always approaches the null
Killing vector on the horizon. What's more, when the black hole is
asymptotically flat and when the coordinate system is
asymptotically static, then the same current also approaches the
time Killing vector at the spatial infinity. We test these results
against various black hole solutions and no exception is found.
The spinor field only needs to satisfy a very general and simple
constraint.

 \newpage

\section{Introduction and Summary of Key Results}

In \cite{mei11} it was noticed that, for the Kerr black hole and
the five dimensional Myers-Perry black hole, there exists a
particular vector field which interpolates between the time
Killing vector at the spatial infinity and the null Killing vector
on the horizon. The existence of such a vector field can be very
interesting in that it may contain important (possibly
non-geometrical) information about the spacetime itself.

In this paper, we want to suggest that the existence of such a
vector is a general feature of all stationary black holes. For all
the examples tested, the vector field always approaches the null
Killing vector on the horizon; when the black hole is
asymptotically flat and when the coordinate system is
asymptotically static, then the same vector field also approaches
the time Killing vector at the spatial infinity. In \cite{mei11},
such features are interpreted as describing a possible fluid flow
underlying the spacetime. Here, we want to leave the physical
interpretation behind and merely demonstrate the existence of the
vector field.

The vector field is constructed using an auxiliary spinor field,
\be\xi^\mu=c_\psi\bpsi\gamma^\mu\psi\,,\quad
\gamma^\mu=e_A^{~~\mu}\gamma^A\,,\label{def.xi}\ee
where $c_\psi$ is a constant and $\gamma^A$ is defined in the
vielbein basis, $e^A=e^A_{~~\mu}dx^\mu$. If $\psi$ was to obey the
Dirac equation, then (\ref{def.xi}) is nothing but the conserved
current of the spinor field. Here we do not require $\psi$ to be a
Dirac fermion, but we will still occasionally refer to
(\ref{def.xi}) as the ``conserved current" for simplicity.

For a stationary and axisymmetric black hole, it is empirically
known that one can always put the metric into one of the following
forms \cite{mei10}
\bea ds^2=-f_t\Delta(dt+f_ad\phi^a)^2 +\frac{f_r}{\Delta} dr^2
+h_i d\theta^{i2}+g_{ab}(d\phi^a-w^adt)(d\phi^b-w^bdt) \,,
\label{metric.general}\\
ds^2=-f_t\Delta(f_ad\phi^a)^2 +\frac{f_r} {\Delta}dr^2 +h_i
d\theta^{i2}+g_{ab}(d\phi^a -w^adt)(d\phi^b -w^bdt)\,,
\label{metric.general2}\eea
where $\Delta=\Delta(r)$, and the functions $f_t, f_a, f_r, h_i,
g_{ab}$ and $w^a$ only depend on $r$ and $\theta^i$'s. The black
hole horizon $r_0$ is located at the (largest) root of
$\Delta(r_0)=0$.  For many solutions, one can explicitly choose
the coordinate system to be non-rotating at the spatial infinity
($r\rightarrow+\infty$), and the spatial coordinates can be
identified with those from a usual spherical coordinate system ---
namely the radius $r$, latitudinal angles $\theta^i$ ($i=1,\cdots,
[\frac{d}2]-1$) and azimuthal angles $\phi^a$
($a=1,\cdots,[\frac{d+1}2]-1$), where $d$ is the dimension of the
spacetime. In such cases, one has \cite{mei10}
\be w^a \rightarrow\Omega^a\quad{\rm as}\quad r\rightarrow r_0\,,
\ee
where $\Omega^a$ is the angular velocity of the black hole in the
$\phi^a$ direction.

It is also empirically known that the functions $f_t, f_r, h_i$
and the matrix $(g_{ab})$ can always be made {\it positive
definite} near the black hole horizon \cite{mei10}. For this
reason, one can always rewrite (\ref{metric.general}) and
(\ref{metric.general2}) in terms of vielbeins,
\be ds_d^2=\eta_{AB}^{~}e^Ae^B\,,\quad A,B=0,\cdots,d-1\,,
\label{metric.vielbein}\ee
where $\eta=diag\{-+\cdots+\}$, and
\bea e^0=\sqrt{f_t\Delta} (dt+f_ad\phi^a)\quad{\rm or}\quad
e^0=\sqrt{f_t\Delta} (f_ad\phi^a)\,,\quad
e^1=\sqrt{\frac{f_r}\Delta} dr\,,\nn\\
e^{1+i}=\sqrt{h_i}d\theta^i ~({\rm no~summation~over~}i)\,:\quad
i=1,\cdots, [\frac{d}2]-1\,,\nn\\
e^{[\frac{d}2]+a}\,:\quad a=1,\cdots,[\frac{d+1}2]-1\,,
\label{def.vielbeins}\eea
where $e^{[\frac{d}2]+a}$'s are obtained by diagonalizing the last
terms in (\ref{metric.general}) and (\ref{metric.general2}). In
general, (\ref{metric.vielbein}) and (\ref{def.vielbeins}) are
only well defined near the black hole horizon, where all the
vielbeins in (\ref{def.vielbeins}) are real. But for many
solutions given by (\ref{metric.general}), the functions $f_t,
f_r, h_i$ and the matrix $(g_{ab})$ are in fact positive definite
in the whole region outside the black hole horizon \cite{mei10}.
So for these solutions, (\ref{metric.vielbein}) and
(\ref{def.vielbeins}) are well defined in the whole region outside
the black hole horizon.

Our main result of this paper is to demonstrate the following two
results:

\begin{description}
\item[Result \#1:] {\it Given (\ref{metric.general}) (or
(\ref{metric.general2})) and that the spinor field $\psi$
obeys\footnote{All the gamma matrices in (\ref{key.constr}) are
defined in the vielbein basis. In even dimensions, $$\gamma_{
even}^d =(-1)^{\frac{d -2}4} \gamma^1 \cdots\gamma^{d-1}
\gamma^0\,,\quad (\gamma_{ even}^d)^\dagger=\gamma_{
even}^d\,,\quad (\gamma_{ even}^d)^2=\one_d\,,$$ and in odd
dimensions,
$$\gamma_{odd}^d =(-1)^{\frac{d -1}4} \gamma^1\cdots \gamma^{d-1}
\gamma^0\propto \one_d\,,\quad (\gamma_{odd}^d)^\dagger= -\gamma_{
odd}^d\,,\quad (\gamma_{odd}^d)^2=-\one_d\,,$$ where $\one_d$ is
the unit matrix in $d$ dimensions.}
\bea (\gamma^0+\gamma^d)\psi=0\quad {\rm or}\quad(\gamma^0
-\gamma^d)\psi=0\,,\label{key.constr}\eea
then (\ref{def.xi}) always reduces to (Note one can always
normalized $\psi$ to let $\xi^t=1$.)
\be\xi^\mu\pd_\mu=\pd_t+w^a\pd_{\phi^a}\,,\label{key.result} \ee
which means
\be\xi^2=\Bigg\{\begin{matrix}-(1+f_aw^a)^2f_t\Delta&:&{\rm for}
(\ref{metric.general})\,,\cr -(f_aw^a)^2f_t\Delta&:&{\rm for}
(\ref{metric.general2})\,,\end{matrix}\ee
vanishes on the horizon. Since $w^a$'s become constants on the
horizon \cite{mei10}, $\xi$ becomes nothing but the null Killing
vector on the horizon.}

\item[Result \#2]{\it When the black hole is asymptotically flat
and when the coordinate system is asymptotically static, then
\be w^a \rightarrow0\quad{\rm as}\quad r\rightarrow
+\infty\,.\label{key.result2}\ee
So in this case, $\xi$ interpolates between the time Killing
vector at the spatial infinity and the null Killing vector on the
horizon.}
\end{description}
Note one can only test (\ref{key.result2}) for cases where one can
explicitly find the vielbeins that are well defined in the whole
region outside the black hole horizon. As a side remark, also note
that given (\ref{metric.general}) (or (\ref{metric.general2})) and
(\ref{key.result}),
\be\nabla_\mu\xi^\mu=0\,,\quad\Longrightarrow\quad
\nabla_\rho\nabla_\mu\xi^\rho=R_{\mu\rho}\xi^\rho\,,\ee
which partially justifies calling $\xi$ the ``conserved current".

In the next section, we will use black hole solutions in various
spacetime dimensions to prove result \#1. Then we will prove
result \#2 in the section that follows. After that, we will
conclude with a short discussion.

\section{Testing Result \#1}

In this section, we will show that (\ref{key.result}) is true for
the general metrics given in (\ref{metric.general}) and
(\ref{metric.general2}). Since both equations in
(\ref{key.constr}) lead to the same result in (\ref{key.result}),
we will only use the first equation of (\ref{key.constr}) in all
the following calculations.

It is difficult to do the calculations for all dimensions at once,
so we will only test the result from three to eight dimensions.
This should give us enough confidence in the generality of the
result.

\subsection{$d=3$}

The gamma matrices in the dreibein basis are chosen
as\footnote{Note (\ref{def.xi}) is independent of the choice of
the gamma matrices.}
\be \gamma^0=i\sigma^3\,, \quad \gamma^1=\sigma^1\,,
\quad\gamma^2=\sigma^2\,,\ee
where $\sigma^{1,2,3}$ are the usual Pauli matrices. The spinor
field is
\be \psi=\left(\begin{matrix}\phi_{1a}+i\phi_{1b}\cr
\phi_{2a}+i\phi_{2b}\end{matrix}\right)\,,\ee
where all the functions are real. From the first equation in
(\ref{key.constr}), we find
\be\phi_{2a}=\phi_{2b}=0\,.\label{psi.3d}\ee

In three dimensions, (\ref{metric.general}) becomes
\be ds^2=-f_t(dt+fd\phi)^2+f_rdr^2 +g_p(d\phi-w dt)^2\,.\ee
The corresponding dreibeins are
\be e^0=\sqrt{f_t}(dt+fd\phi)\,,\quad e^1=\sqrt{f_r}dr\,,\quad
e^2=\sqrt{g_p}(d\phi-w dt)\,.\ee
Plug (\ref{psi.3d}) into (\ref{def.xi}), we find
\bea \xi^\mu=\pd_t+w\pd_\phi\,,\eea
just as given in (\ref{key.result}).

Similarly, (\ref{metric.general2}) becomes
\be ds^2=-f_t(fd\phi)^2+f_rdr^2 +g_p(d\phi-w dt)^2\,.\ee
The corresponding dreibeins are
\be e^0=\sqrt{f_t}(fd\phi)\,,\quad e^1=\sqrt{f_r}dr\,,\quad
e^2=\sqrt{g_p}(d\phi-w dt)\,.\ee
Plug (\ref{psi.3d}) into (\ref{def.xi}), we also get
\be \xi^\mu\pd_\mu=\pd_t+w\pd_\phi\,,\ee
as given in (\ref{key.result}).

\subsection{d=4}

The gamma matrices in the vierbein basis are chose as
\be \gamma^0=i\sigma^3\otimes\one_2\,,\quad
\gamma^j=-\sigma^2\otimes\sigma^j\,,\quad j=1,2,3\,.\ee
The spinor field is
\be \psi=\left(\begin{matrix}\phi_{1a}+i\phi_{1b}\cr
\phi_{2a}+i\phi_{2b}\cr \phi_{3a}+i\phi_{3b}\cr
\phi_{4a}+i\phi_{4b}\end{matrix}\right)\,,\ee
where all the functions are real. From the first equation in
(\ref{key.constr}), we find
\be\phi_{1a}=\phi_{3b}\,,\quad \phi_{1b}=-\phi_{3a}\,,\quad
\phi_{2a}=\phi_{4b}\,,\quad \phi_{2b}=-\phi_{4a}\,.
\label{psi.4d}\ee

In four dimensions, (\ref{metric.general}) becomes
\be ds^2=-f_t(dt+fd\phi)^2+f_rdr^2 +f_y d\theta^2 +g_p(d\phi-w
dt)^2\,.\ee
The corresponding vierbeins are
\be e^0=\sqrt{f_t}(dt+fd\phi)\,,\quad e^1=\sqrt{f_r}dr\,,\quad
e^2=\sqrt{f_y}d\theta\,,\quad e^3=\sqrt{g_p}(d\phi-w dt)\,.\ee
Plug (\ref{psi.4d}) into (\ref{def.xi}), we find
\bea \xi^\mu=\pd_t+w\pd_\phi\,,\eea
just as given in (\ref{key.result}).

Similarly, (\ref{metric.general}) becomes
\be ds^2=-f_t(fd\phi)^2+f_rdr^2 +f_y d\theta^2 +g_p(d\phi-w
dt)^2\,.\ee
The corresponding vierbeins are
\be e^0=\sqrt{f_t}(fd\phi)\,,\quad e^1=\sqrt{f_r}dr\,,\quad
e^2=\sqrt{f_y}d\theta\,,\quad e^3=\sqrt{g_p}(d\phi-w dt)\,.\ee
Plug (\ref{psi.4d}) into (\ref{def.xi}), we find
\bea \xi^\mu=\pd_t+w\pd_\phi\,,\eea
also as given in (\ref{key.result}).

\subsection{d=5}

The gamma matrices in the fuenfbein basis is taken to be
\be \gamma^0=i\sigma^1\otimes\one_2\,,\quad
\gamma^4=\sigma^3\otimes\one_2\,,\quad
\gamma^j=-\sigma^2\otimes\sigma^j\,,\quad j=1,2,3\,.\ee
The spinor field is
\be \psi=\left(\begin{matrix}\phi_{1a}+i\phi_{1b}\cr
\phi_{2a}+i\phi_{2b}\cr \phi_{3a}+i\phi_{3b}\cr
\phi_{4a}+i\phi_{4b}\end{matrix}\right)\,,\ee
where all the functions are real. From the first equation in
(\ref{key.constr}), we find
\be\phi_{1a}=\phi_{3a}\,,\quad \phi_{1b}=\phi_{3b}\,,\quad
\phi_{2a}=\phi_{4a}\,,\quad \phi_{2b}=\phi_{4b}\,.
\label{psi.5d}\ee

In five dimensions, (\ref{metric.general}) becomes
\bea ds^2=-f_t(dt+f_1d\phi_1+f_2d\phi_2)^2+f_rdr^2 +f_y
d\theta^2+g_{22}(d\phi_2-w_2 dt)^2\nn\\
+g_{11}\Big[d\phi_1-w_1 dt+g_{12}(d\phi_2-w_2 dt)\Big]^2 \,.\eea
The corresponding fuenfbeins are
\bea e^0=\sqrt{f_t}(dt+f_1d\phi_1+f_2d\phi_2)\,,\quad
e^1=\sqrt{f_r}dr\,,\quad e^2=\sqrt{f_y}d\theta\,,\nn\\
e^3=\sqrt{g_{11}}\Big[d\phi_1-w_1 dt+g_{12}(d\phi_2-w_2
dt)\Big]\,,\quad e^4=\sqrt{g_{22}}(d\phi_2-w_2 dt)\,.\eea
Plug (\ref{psi.5d}) into (\ref{def.xi}), we find
\bea \xi^\mu=\pd_t+w_1\pd_{\phi_1}+w_2\pd_{\phi_2}\,,\eea
just as given in (\ref{key.result}).

Similarly, (\ref{metric.general2}) becomes
\bea ds^2=-f_t(f_1d\phi_1+f_2d\phi_2)^2+f_rdr^2 +f_y
d\theta^2+g_{22}(d\phi_2-w_2 dt)^2\nn\\
+g_{11}\Big[d\phi_1-w_1 dt+g_{12}(d\phi_2-w_2 dt)\Big]^2 \,.\eea
The corresponding fuenfbeins are
\bea e^0=\sqrt{f_t}(f_1d\phi_1+f_2d\phi_2)\,,\quad
e^1=\sqrt{f_r}dr\,,\quad e^2=\sqrt{f_y}d\theta\,,\nn\\
e^3=\sqrt{g_{11}}\Big[d\phi_1-w_1 dt+g_{12}(d\phi_2-w_2
dt)\Big]\,,\quad e^4=\sqrt{g_{22}}(d\phi_2-w_2 dt)\,.\eea
Plug (\ref{psi.5d}) into (\ref{def.xi}), we find
\bea \xi^\mu=\pd_t+w_1\pd_{\phi_1}+w_2\pd_{\phi_2}\,,\eea
also as given in (\ref{key.result}).

\subsection{d=6}

The gamma matrices in the sechsbein basis are taken to be
\be \gamma^0=i\sigma^1\otimes\one_4\,,\quad
\gamma^5=\sigma^3\otimes\one_4\,,\quad
\gamma^j=-\sigma^2\otimes\gamma_4^j\,,\quad j=1,2,3,4\,,\ee
where $\gamma_4^{1,2,3,4}$ are gamma matrices in the vierbein
basis from four-dimensions, and we have replaced $\gamma_4^0$ by
$\gamma_4^4=-i\gamma_4^0$. The spinor field is
\be \psi=\left(\begin{matrix}\phi_{1a}+i\phi_{1b}\cr\vdots\cr
\phi_{8a}+i\phi_{8b}\end{matrix}\right)\,,\ee
where all the functions are real. From the first equation in
(\ref{key.constr}), we find
\bea\phi_{1a}=-\phi_{3a}\,,\quad \phi_{1b}=-\phi_{3b}\,,\quad
\phi_{2a}=-\phi_{4a}\,,\quad \phi_{2b}=-\phi_{4b}\,,\nn\\
\phi_{5a}=\phi_{7a}\,,\quad \phi_{5b}=\phi_{7b}\,,\quad
\phi_{6a}=\phi_{8a}\,,\quad \phi_{6b}=\phi_{8b}\,.
\label{psi.6d}\eea

In six dimensions, (\ref{metric.general}) becomes
\bea ds^2=-f_t(dt+f_1d\phi_1+f_2d\phi_2)^2+f_rdr^2 +f_{11}
d\theta_1^2+f_{22}d\theta_2^2\nn\\
+g_{11}\Big[d\phi_1-w_1 dt+g_{12}(d\phi_2-w_2 dt)\Big]^2
+g_{22}(d\phi_2-w_2 dt)^2 \,.\eea
The corresponding sechsbeins are
\bea e^0=\sqrt{f_t}(dt+f_1d\phi_1+f_2d\phi_2)\,,\quad
e^1=\sqrt{f_r}dr\,,\quad e^2=\sqrt{f_{11}}d\theta_1
\,,\quad e^3=\sqrt{f_{22}}d\theta_2\,,\nn\\
e^4=\sqrt{g_{11}}\Big[d\phi_1-w_1 dt+g_{12}(d\phi_2-w_2
dt)\Big]\,,\quad e^5=\sqrt{g_{22}}(d\phi_2-w_2 dt)\,.\eea
Plug (\ref{psi.6d}) into (\ref{def.xi}), we find
\bea \xi^\mu=\pd_t+w_1\pd_{\phi_1}+w_2\pd_{\phi_2}\,,\eea
just as given in (\ref{key.result}).

Similarly, (\ref{metric.general2}) becomes
\bea ds^2=-f_t(f_1d\phi_1+f_2d\phi_2)^2+f_rdr^2 +f_{11}
d\theta_1^2+f_{22}d\theta_2^2\nn\\
+g_{11}\Big[d\phi_1-w_1 dt+g_{12}(d\phi_2-w_2 dt)\Big]^2
+g_{22}(d\phi_2-w_2 dt)^2 \,.\eea
The corresponding sechsbeins are
\bea e^0=\sqrt{f_t}(f_1d\phi_1+f_2d\phi_2)\,,\quad
e^1=\sqrt{f_r}dr\,,\quad e^2=\sqrt{f_{11}}d\theta_1
\,,\quad e^3=\sqrt{f_{22}}d\theta_2\,,\nn\\
e^4=\sqrt{g_{11}}\Big[d\phi_1-w_1 dt+g_{12}(d\phi_2-w_2
dt)\Big]\,,\quad e^5=\sqrt{g_{22}}(d\phi_2-w_2 dt)\,.\eea
Plug (\ref{psi.6d}) into (\ref{def.xi}), we find
\bea \xi^\mu=\pd_t+w_1\pd_{\phi_1}+w_2\pd_{\phi_2}\,,\eea
also as given in (\ref{key.result}).

\subsection{d=7}

The gamma matrices in the siebbein basis are taken to be
\be \gamma^0=i\sigma^1\otimes\one_5\,,\quad
\gamma^6=\sigma^3\otimes\one_5\,,\quad
\gamma^j=-\sigma^2\otimes\gamma_5^j\,,\quad j=1,2,3,4,5\,,\ee
where $\gamma_5^{1,2,3,4,5}$ are gamma matrices in the fuenfbein
basis from five-dimensions, and we have replaced $\gamma_5^0$ by
$\gamma_5^5=-i\gamma_5^0$. The spinor field is
\be \psi=\left(\begin{matrix}\phi_{1a}+i\phi_{1b}\cr\vdots\cr
\phi_{8a}+i\phi_{8b}\end{matrix}\right)\,,\ee
where all the functions are real. From the first equation in
(\ref{key.constr}), we find
\bea\phi_{1a}=\phi_{5a}\,,\quad \phi_{1b}=\phi_{5b}\,,\quad
\phi_{2a}=\phi_{6a}\,,\quad \phi_{2b}=\phi_{6b}\,,\nn\\
\phi_{3a}=\phi_{7a}\,,\quad \phi_{3b}=\phi_{7b}\,,\quad
\phi_{4a}=\phi_{8a}\,,\quad \phi_{4b}=\phi_{8b}\,.
\label{psi.7d}\eea

In seven dimensions, (\ref{metric.general}) becomes
\bea ds^2=-f_t(dt+f_1d\phi_1+f_2d\phi_2+f_3d\phi_3)^2+f_rdr^2
+f_{11}d\theta_1^2+f_{22}d\theta_2^2\nn\\
+g_{11}\Big[d\phi_1-w_1 dt+g_{12}(d\phi_2-w_2 dt)
+g_{13}(d\phi_3-w_3 dt)\Big]^2\nn\\
+g_{22}\Big[d\phi_2-w_2 dt+g_{23}(d\phi_3-w_3 dt)\Big]^2
+g_{33}(d\phi_3-w_3 dt)^2 \,.\eea
The corresponding siebbeins are
\bea e^0=\sqrt{f_t}(dt+f_1d\phi_1+f_2d\phi_2+f_3d\phi_3)\,,\quad
e^1=\sqrt{f_r}dr\,,\quad e^2=\sqrt{f_{11}}d\theta_1\,,\nn\\
e^3=\sqrt{f_{22}}d\theta_2\,,\quad e^4=\sqrt{g_{11}} \Big[d\phi_1
-w_1 dt+g_{12}(d\phi_2-w_2 dt)+g_{13}(d\phi_3-w_3 dt)\Big]\,,\nn\\
e^5=\sqrt{g_{22}}\Big[d\phi_2-w_2 dt+g_{23}(d\phi_3-w_3
dt)\Big]\,,\quad e^6=\sqrt{g_{33}}(d\phi_3-w_3 dt)\,.\eea
Plug (\ref{psi.7d}) into (\ref{def.xi}), we find
\bea \xi^\mu=\pd_t+w_1\pd_{\phi_1}+w_2\pd_{\phi_2}
+w_3\pd_{\phi_3}\,,\eea
just as given in (\ref{key.result}).

Similarly, (\ref{metric.general2}) becomes
\bea ds^2=-f_t(f_1d\phi_1+f_2d\phi_2+f_3d\phi_3)^2+f_rdr^2
+f_{11}d\theta_1^2+f_{22}d\theta_2^2\nn\\
+g_{11}\Big[d\phi_1-w_1 dt+g_{12}(d\phi_2-w_2 dt)
+g_{13}(d\phi_3-w_3 dt)\Big]^2\nn\\
+g_{22}\Big[d\phi_2-w_2 dt+g_{23}(d\phi_3-w_3 dt)\Big]^2
+g_{33}(d\phi_3-w_3 dt)^2 \,.\eea
The corresponding siebbeins are
\bea e^0=\sqrt{f_t}(f_1d\phi_1+f_2d\phi_2+f_3d\phi_3)\,,\quad
e^1=\sqrt{f_r}dr\,,\quad e^2=\sqrt{f_{11}}d\theta_1\,,\nn\\
e^3=\sqrt{f_{22}}d\theta_2\,,\quad e^4=\sqrt{g_{11}} \Big[d\phi_1
-w_1 dt+g_{12}(d\phi_2-w_2 dt)+g_{13}(d\phi_3-w_3 dt)\Big]\,,\nn\\
e^5=\sqrt{g_{22}}\Big[d\phi_2-w_2 dt+g_{23}(d\phi_3-w_3
dt)\Big]\,,\quad e^6=\sqrt{g_{33}}(d\phi_3-w_3 dt)\,.\eea
Plug (\ref{psi.7d}) into (\ref{def.xi}), we find
\bea \xi^\mu=\pd_t+w_1\pd_{\phi_1}+w_2\pd_{\phi_2}
+w_3\pd_{\phi_3}\,,\eea
also as given in (\ref{key.result}).

\subsection{d=8}

The gamma matrices in the vielbein basis are taken to be
\be \gamma^0=i\sigma^1\otimes\one_6\,,\quad
\gamma^7=\sigma^3\otimes\one_6\,,\quad
\gamma^j=-\sigma^2\otimes\gamma_6^j\,,\quad j=1,2,3,4,5,6\,,\ee
where $\gamma_6^{1,2,3,4,5,6}$ are gamma matrices in the sechsbein
basis from six-dimensions, and we have replaced $\gamma_6^0$ by
$\gamma_6^6=-i\gamma_6^0$. The spinor field is
\be \psi=\left(\begin{matrix}\phi_{1a}+i\phi_{1b}\cr\vdots\cr
\phi_{16a}+i\phi_{16b}\end{matrix}\right)\,,\ee
where all the functions are real. From the first equation in
(\ref{key.constr}), we find
\bea\phi_{1a}=\phi_{7b}\,,\quad \phi_{1b}=-\phi_{7a}\,,\quad
\phi_{2a}=\phi_{8b}\,,\quad \phi_{2b}=-\phi_{8a}\,,\nn\\
\phi_{3a}=\phi_{5b}\,,\quad \phi_{3b}=-\phi_{5a}\,,\quad
\phi_{4a}=\phi_{6b}\,,\quad \phi_{4b}=-\phi_{6a}\,,\nn\\
\phi_{9a}=-\phi_{15b}\,,\quad \phi_{9b}=\phi_{15a}\,,\quad
\phi_{10a}=-\phi_{16b}\,,\quad \phi_{10b}=\phi_{16a}\,,\nn\\
\phi_{11a}=-\phi_{13b}\,,\quad \phi_{11b}=\phi_{13a}\,,\quad
\phi_{12a}=-\phi_{14b}\,,\quad \phi_{12b}=\phi_{14a} \,.
\label{psi.8d}\eea

In eight dimensions, (\ref{metric.general}) becomes
\bea ds^2=-f_t(dt+f_1d\phi_1+f_2d\phi_2+f_3d\phi_3)^2+f_rdr^2
+f_{11}d\theta_1^2+f_{22}d\theta_2^2 +f_{33}d\theta_3^2\nn\\
+g_{11}\Big[d\phi_1-w_1 dt+g_{12}(d\phi_2-w_2 dt)
+g_{13}(d\phi_3-w_3 dt)\Big]^2\nn\\
+g_{22}\Big[d\phi_2-w_2 dt+g_{23}(d\phi_3-w_3 dt)\Big]^2
+g_{33}(d\phi_3-w_3 dt)^2 \,.\label{metric.8d}\eea
The corresponding vielbeins are
\bea e^0=\sqrt{f_t}(dt+f_1d\phi_1+f_2d\phi_2+f_3d\phi_3)\,,\quad
e^1=\sqrt{f_r}dr\,,\nn\\
e^2=\sqrt{f_{11}}d\theta_1\,,\quad
e^3=\sqrt{f_{22}}d\theta_2\,,\quad
e^4=\sqrt{f_{33}}d\theta_3\,,\nn\\
e^5=\sqrt{g_{11}} \Big[d\phi_1-w_1 dt+g_{12}(d\phi_2-w_2 dt)
+g_{13}(d\phi_3-w_3 dt)\Big]\,,\nn\\
e^6=\sqrt{g_{22}}\Big[d\phi_2-w_2 dt+g_{23}(d\phi_3-w_3
dt)\Big]\,,\quad e^7=\sqrt{g_{33}}(d\phi_3-w_3 dt)\,.\eea
Plug (\ref{psi.7d}) into (\ref{def.xi}), we find
\bea \xi^\mu=\pd_t+w_1\pd_{\phi_1}+w_2\pd_{\phi_2}
+w_3\pd_{\phi_3}\,,\eea
just as given in (\ref{key.result}).

Similarly, (\ref{metric.general2}) becomes
\bea ds^2=-f_t(f_1d\phi_1+f_2d\phi_2+f_3d\phi_3)^2+f_rdr^2
+f_{11}d\theta_1^2+f_{22}d\theta_2^2 +f_{33}d\theta_3^2\nn\\
+g_{11}\Big[d\phi_1-w_1 dt+g_{12}(d\phi_2-w_2 dt)
+g_{13}(d\phi_3-w_3 dt)\Big]^2\nn\\
+g_{22}\Big[d\phi_2-w_2 dt+g_{23}(d\phi_3-w_3 dt)\Big]^2
+g_{33}(d\phi_3-w_3 dt)^2 \,.\eea
The corresponding vielbeins are
\bea e^0=\sqrt{f_t}(f_1d\phi_1+f_2d\phi_2+f_3d\phi_3)\,,\quad
e^1=\sqrt{f_r}dr\,,\nn\\
e^2=\sqrt{f_{11}}d\theta_1\,,\quad
e^3=\sqrt{f_{22}}d\theta_2\,,\quad
e^4=\sqrt{f_{33}}d\theta_3\,,\nn\\
e^5=\sqrt{g_{11}} \Big[d\phi_1-w_1 dt+g_{12}(d\phi_2-w_2 dt)
+g_{13}(d\phi_3-w_3 dt)\Big]\,,\nn\\
e^6=\sqrt{g_{22}}\Big[d\phi_2-w_2 dt+g_{23}(d\phi_3-w_3
dt)\Big]\,,\quad e^7=\sqrt{g_{33}}(d\phi_3-w_3 dt)\,.\eea
Plug (\ref{psi.7d}) into (\ref{def.xi}), we find
\bea \xi^\mu=\pd_t+w_1\pd_{\phi_1}+w_2\pd_{\phi_2}
+w_3\pd_{\phi_3}\,,\eea
also as given in (\ref{key.result}).

\subsection{Summary of the section}

To summarize this section, we have explicitly shown that
(\ref{key.result}) is true for any metric of the form
(\ref{metric.general}) or (\ref{metric.general2}), given that
(\ref{key.constr}) is satisfied. The calculation is only done in
three through eight dimensions. But we do not see any particular
reason that such a pattern will break down in higher dimensions.
Given the fact that (\ref{metric.general}) and
(\ref{metric.general2}) are quite general structures for all known
stationary and axisymmetric black holes \cite{mei10}, one can
conclude that result \#1 holds for all such black holes.

\section{Testing Result \#2}

In this section, we will use examples in different dimensions to
demonstrate result \#2. In this case, we need to make sure that
the coordinate system is non-rotating at the spatial infinity and
that the vielbeins (\ref{def.vielbeins}) are well defined in the
whole region outside the black hole horizon. So more detail of
each specific solution will be needed.

\subsection{$d=3$}

In three spacetime dimensions, an interesting example is the BTZ
black hole \cite{btz92},
\be ds^2=-fd\hat{t}^2+\frac{dr^2}f+r^2\Big(d\hat\phi-\frac{J}{2
r^2}d\hat{t}\Big)^2\,,\quad f=-m+g^2r^2+\frac{J^2}{4r^2}\,,
\label{metric.btz}\ee
which solves the Einstein equation with a cosmological constant,
\be R_{\mu\nu}=\frac{2\Lambda}{d-2}g_{\mu\nu}\,,\quad
\Lambda=-\frac{(d-1)(d-2)}2g^2\,.\label{eom.Lambda}\ee
The coordinates in (\ref{metric.btz}) are related to the static
ones by
\bea d\hat{t}=\frac1{\sqrt{2(m-\td{m})}}\Big(\frac{J g}{
\sqrt{\td{m}}}dt +\frac{\sqrt{\td{m}}}gd\phi\Big)\,,\nn\\
d\hat\phi=\frac1{\sqrt{2(m-\td{m})}}\Big(\frac{J g}{
\sqrt{\td{m}}}d\phi +g\sqrt{\td{m}}dt\Big)\,,\eea
where $\td{m}=m-\sqrt{m^2-J^2g^2}$. Now the metric becomes
\be ds^2=-\frac{\Delta r_0^2}{r_0^2-r_c^2}\Big(dt+\frac{r_c}{gr_0}
d\phi\Big)+\frac{dr^2}{\Delta}+\frac{r_0^2(r^2-r_c^2)^2}{r^2
(r_0^2-r_c^2)}\Big[d\phi+\frac{gr_c(r^2-r_0^2)}{r_0(r^2-r_c^2)}
dt\Big]^2\,.\label{metric.btz2}\ee
where
\be\Delta=\frac{g^2(r^2-r_0^2)(r^2-r_c^2)}{r^2}\,,\quad
r_c=\frac{J}{2gr_0}<r_0\,.\ee
For this metric, the horizon angular velocity is zero, $\Omega=0$.

The dreibeins can be read off (\ref{metric.btz2}) in a
straightforward manor. From (\ref{def.xi}) and (\ref{psi.3d}), we
find
\be \xi^\mu\pd_\mu=\pd_t-\frac{gr_c(r^2-r_0^2)}{r_0
(r^2-r_c^2)}\pd_\phi\,,\ee
just as given in (\ref{key.result}). As $r\rightarrow+\infty$,
\be \xi\rightarrow\pd_t-\frac{gr_c}{r_0}\pd_\phi\,.\ee
For the BTZ black hole, one can never set $g=0$, so $\xi$ will not
become the time Killing vector at the spatial infinity. On the
other hand, one sees that $\xi=\pd_t$ on the horizon when $r=r_0$.
This is a peculiar feature of the BTZ black hole.

\subsection{d=4}

In four dimensions, we consider the rotating solution in U(1)$^4$
gauged supergravity with four charges pairwise equal
\cite{cclp04a}. The metric is given by (\ref{metric.vielbein})
with the vierbeins,
\bea e^0=\sqrt{\frac{R}{H(r^2+y^2)}}\left[d\hat{t}
-\frac{a^2-y^2}{a(1-g^2a^2)}d\hat\phi\right]\,,\nn\\
e^1=\sqrt{\frac{H(r^2+y^2)}{R}}dr\,,\quad
e^2=\sqrt{\frac{H(r^2+y^2)}{Y}}dy\,,\nn\\
e^3=\sqrt{\frac{Y}{H(r^2+y^2)}}\frac{r_1 r_2+a^2}{a
(1-g^2a^2)}\left[d\hat\phi -\frac{a(1-g^2a^2)}{r_1
r_2+a^2}d\hat{t}\right]\,,\eea
where $r_1=r+2ms_1^2$, $r_2=r+2ms_2^2$ and
\bea R=r^2+a^2-2mr+g^2r_1r_2[r_1r_2+a^2]\,,\nn\\
Y=(1-g^2y^2)(a^2-y^2)\,,\quad H=\frac{r_1r_2 +y^2}{r^2+y^2}\,.\eea
The coordinates are related to the static ones by
\be d\hat{t}=dt\,,\quad d\hat\phi=d\phi-g^2adt\,.\ee
The horizon is located at $R(r_0)=0$, and the angular velocity is
\be \Omega=\frac{a(1+g^2r_{10}r_{20})}{r_{10}r_{20}+a^2}\,,\ee
where $r_{10}=r_1(r_0)$ and $r_{20}=r_2(r_0)$.

From (\ref{def.xi}) and (\ref{psi.4d}), we find
\be \xi^\mu\pd_\mu=\pd_t+\frac{a(1+g^2r_1r_2)}{r_1r_2
+a^2}\pd_\phi\,,\ee
just as given in (\ref{key.result}). We see that
$\xi\rightarrow\pd_t+g^2a\pd_\phi$ as $r\rightarrow+\infty$. The
solution is asymptotically flat when $g=0$. In this case,
$\xi\rightarrow\pd_t$ as $r\rightarrow+\infty$.

It will also be interesting to consider the single-charge and
two-charge rotating black hole in the gauged supergravity
\cite{chow10a,chow10b}, and the four-charge black hole in the
ungauged supergravity \cite{cvetic.youm96.4d,cclp04a}. But for
these solutions, we have not been able to put the metrics into
desired the form. So the corresponding calculation is not done.

\subsection{d=5}

In five dimensions, we consider the rotating solution in $U(1)^3$
gauged supergravity with two of the charges equal
\cite{mei.pope07}. The metric is given by (\ref{metric.vielbein})
with the fuenfbeins,
\bea e^0=\frac{h_3^{1/6}r\sqrt{X}}{h_1^{2/3}
\sqrt{r_3}}\sigma_t\,,\quad
e^1=\frac{h_1^{1/3}h_3^{1/6}}{\sqrt{X}}dr\,,\quad
e^2=\frac{h_1^{1/3}h_3^{1/6}}{\sqrt{1-g^2y^2}}d\theta\,,\nn\\
e^3=\frac{(a^2-b^2)h_3^{1/6}\sqrt{1-g^2y^2}\cos\theta
\sin\theta}{h_1^{2/3}y}\sigma_a\,,\quad
e^4=\frac{(abh_3+2mc_3s_3y^2)\sigma_t +h_1r_3
\sigma_b}{h_1^{2/3}h_3^{1/3}y\sqrt{r_3}}\,,\eea
where
\bea\sigma_t=\frac{1-g^2y^2}{Z_aZ_b}dt-\frac{a\sin^2\theta}{Z_a}
d\phi_1 -\frac{b\cos^2\theta}{Z_b}d\phi_2\,,\nn\\
\sigma_a=\frac{1+g^2r_1}{Z_aZ_b}dt-\frac{a(a^2+r_1)}{Z_a(a^2
-b^2)}d\phi_1 +\frac{b(b^2+r_1)}{Z_b(a^2-b^2)}d\phi_2\,,\nn\\
\sigma_b=\frac{g^2ab(1-g^2y^2)}{Z_aZ_b}dt-\frac{b\sin^2\theta}{Z_a}
d\phi_1 -\frac{a\cos^2\theta}{Z_b}d\phi_2\,,\nn\\
y^2=a^2\cos^2\theta+b^2\sin^2\theta\,,\quad Z_a=1-g^2a^2\,,\quad
Z_b=1-g^2b^2\,,\nn\\
r_1=r_2+2ms_1^2\,,\quad r_3=r_2+2ms_3^2\,,\quad
r_2=r^2-\frac23m(2s_1^2+s_3^2)\,,\nn\\
h_1=r_1+y^2\,,\quad h_3=r_3+y^2\,,\quad c_1=\sqrt{1+s_1^2}\,,\quad
c_3=\sqrt{1+s_3^2}\,,\nn\\
X=r^{-2}\Big[(r_2+a^2)(r_2+b^2)+g^2(r_1+a^2)(r_1+b^2)r_3\nn\\
-2m(r_2-2abc_3s_3-a^2s_3^2-b^2s_3^2)\Big]\,.\eea
The gauge fields are
\be A_1=A_2=\frac{2mc_1s_1}{h_1}\sigma_t\,,\quad
A_3=\frac{2m[c_3s_3\sigma_t-(s_1^2-s_3^2)\sigma_b]}{h_3}\,.\ee
The horizon is located at $X(r_0)=0$, and the angular velocities
are
\bea\Omega_a=\frac{b(a b+2ms_3c_3)+a[1+g^2(r_{10}+b^2)] r_{30}}{a
b(a b+2ms_3c_3)+(r_{10}+a^2+b^2)r_{30}}\,,\nn\\
\Omega_b=\frac{a(a b+2ms_3c_3)+b[1+g^2(r_{10}+a^2)] r_{30}}{a b(a
b+2ms_3c_3)+(r_{10}+a^2+b^2)r_{30}}\,,\eea
where $r_{10}=r_1(r=r_0)$ and $r_{30}=r_3(r=r_0)$.

From (\ref{def.xi}) and (\ref{psi.5d}), we find
\be \xi^\mu\pd_\mu=\pd_t +w_1\pd_{\phi_1} +w_2\pd_{\phi_2}\,,\ee
with
\bea w_1=\frac{b(a b+2ms_3c_3)+a[1+g^2(r_1+b^2)] r_3}{a b (a b
+2ms_3c_3)+(r_1+a^2+b^2)r_3}\,,\nn\\
w_2=\frac{a(a b+2ms_3c_3)+b[1+g^2(r_1+a^2)] r_3}{a b (a b
+2ms_3c_3)+(r_1+a^2+b^2)r_3}\,.\eea
It is obvious that the result agrees with (\ref{key.result}). In
the limit $r\rightarrow+\infty$, we find
\be w_1=g^2a+\cO(\frac1r)\,,\quad w_2=g^2b+\cO(\frac1r)\,.\ee
In the case $g=0$, the solution is asymptotically flat, and we
have $\xi=\pd_t$ as $r\rightarrow+\infty$.

It will also be interesting to consider the equal rotation
solution in $U(1)^3$ gauged supergravity with arbitrary charges
\cite{clp04b} and the Cveti\v c-Youm solution \cite{cvetic.youm96}
in ungauged supergravity. But for these solutions, we have not
been able to put the metrics into the desired form. So the
corresponding calculation is not done.

\subsection{d=6}

In six dimensions, we consider the single-charge two-rotation
solution in $SU(2)$ gauged supergravity found in \cite{chow08}.
The metric is given in (\ref{metric.vielbein}) with the
sechsbeins,
\bea e^0=\sqrt{\frac{R}{H^{3/2}U}}\cA\,,\quad
e^1=\sqrt{\frac{H^{1/2}U}R}dr\,,\nn\\
e^2=\sqrt{\frac{H^{1/2}(r^2+y^2)(y^2-z^2)}Y}dy\,,\quad
e^3=\sqrt{\frac{H^{1/2}(r^2+z^2)(z^2-y^2)}Z}dz\,,\nn\\
e^4=\sqrt{\frac{H^{1/2}Y}{(r^2+y^2)(y^2-z^2)}}\cA_Y\,,\quad
e^5=\sqrt{\frac{H^{1/2}Z}{(r^2+z^2)(z^2-y^2)}}\cA_Z\,,\eea
where (Note $\hat\phi_1=\phi_1-g^2at$ and
$\hat\phi_2=\phi_2-g^2bt$)
\bea\cA_Y=dt-(r^2+a^2)(a^2-z^2) \frac{d\hat{
\phi}_1}{\epsilon_1}-(r^2+b^2)(b^2-z^2) \frac{d\hat{
\phi}_2}{\epsilon_2}-\frac{qr\wtd\cA}{H U}\,,\nn\\
\cA_Z=dt-(r^2+a^2)(a^2-y^2)\frac{d\hat{
\phi}_1}{\epsilon_1}-(r^2+b^2)(b^2-y^2) \frac{d\hat{
\phi}_2}{\epsilon_2}-\frac{qr\wtd\cA}{H U}\,,\nn\\
\cA=dt-(a^2-y^2)(a^2-z^2)\frac{d\hat{\phi}_1}{\epsilon_1}
-(b^2-y^2)(b^2-z^2)\frac{d\hat{\phi}_2}{\epsilon_2}\,,\nn\\
R=(r^2+a^2)(r^2+b^2)+g^2[r(r^2+a^2)+q][r(r^2+b^2)+q]-2mr\,.\eea
More detail of the solution can be found in
\cite{chow08,chow.cvetic.lu.pope08}.

From (\ref{def.xi}) and (\ref{psi.6d}), we find that
\be \xi^\mu\pd_\mu=\pd_t +w_1\pd_{\phi_1} +w_2\pd_{\phi_2}\,,\ee
with
\bea w_1=\frac{a[g^2qr+(b^2+r^2)(1+g^2r^2)]}{q
r+(a^2+r^2)(b^2+r^2)}\,,\nn\\
w_2=\frac{b[g^2qr+(a^2+r^2)(1+g^2r^2)]}{q
r+(a^2+r^2)(b^2+r^2)}\,.\eea
Note $\Omega_a=w_1(r_0)$ and $\Omega_b=w_2(r_0)$ are just the two
angular velocities of the black hole. It is obvious that the
result agrees with (\ref{key.result}). In the limit
$r\rightarrow+\infty$, we find
\be w_1=g^2a+\cO(\frac1r)\,,\quad w_2=g^2b+\cO(\frac1r)\,.\ee
In the case $g=0$, the solution is asymptotically flat, and we
have $\xi=\pd_t$ as $r\rightarrow+\infty$.

\subsection{d=7}

In seven dimensions, we consider the single-charge three-rotation
solution in $SO(5)$ gauged supergravity found in \cite{chow07}.
The metric is given in (\ref{metric.vielbein}) with the siebbeins,
\bea e^0=\sqrt{\frac{R}{H^{8/5}U}}\cA\,,\quad
e^1=\sqrt{\frac{H^{2/5}U}R}dr\,,\nn\\
e^2=\sqrt{\frac{H^{2/5}(r^2+y^2)(y^2-z^2)}Y}dy\,,\quad
e^3=\sqrt{\frac{H^{2/5}(r^2+z^2)(z^2-y^2)}Z}dz\,,\nn\\
e^4=\sqrt{\frac{H^{2/5}Y}{(r^2+y^2)(y^2-z^2)}}\cA_Y\,,\quad
e^5=\sqrt{\frac{H^{2/5}Z}{(r^2+z^2)(z^2-y^2)}}\cA_Z\,,\quad
e^6=\frac{a_1a_2a_3}{r y z}\cA_7\,,\eea
where (Note $\hat\phi_i=\phi_i-g^2a_it$, $i=1,2,3$.)
\bea\cA_Y=dt-\sum_{i=1}^3\frac{(\hat{r}^2+a_i^2)\gamma_i}{a_i^2
-y^2} \frac{d \hat{\phi}_i}{\epsilon_i} -\frac{q}{H U}\cA\,,\quad
\cA_Z=dt-\sum_{i=1}^3\frac{(\hat{r}^2+a_i^2)\gamma_i}{a_i^2
-z^2}\frac{d\hat{\phi}_i}{\epsilon_i}-\frac{q}{H U}\cA\,,\nn\\
\cA_7=dt-\sum_{i=1}^3\frac{(\hat{r}^2+a_i^2)\gamma_i}{a_i^2}
\frac{d \hat{\phi}_i}{\epsilon_i}-\frac{q}{H U}\Big(1+\frac{gy^2
z^2}{a_1a_2a_3}\Big)\cA\,,\quad \cA=d\hat{t}-\sum_{i=1}^3\gamma_i
\frac{d \hat \phi_i}{\epsilon_i}\,,\nn\\
R=\frac{1+g^2r^2}{r^2}\prod_{i=1}^3(r^2+a_i^2)+qg^2(2r^2+a_1^2
+a_2^2+a_3^2)-\frac{2qga_1a_2a_3}{r^2}+\frac{q^2g^2}{r^2}-2m
\,.\eea
More detail of the solution can be found in
\cite{chow.cvetic.lu.pope08,chow07}.

From (\ref{def.xi}) and (\ref{psi.7d}), we find
\be \xi^\mu\pd_\mu=\pd_t +w_1\pd_{\phi_1} +w_2\pd_{\phi_2}
+w_3\pd_{\phi_3}\,,\ee
with
\bea w_1=\frac{a_1(r^2+a_2^2)(r^2+a_3^2)(1+g^2r^2)-g q(a_2a_3-a_1
g r^2)}{(r^2+a_1^2)(r^2+a_2^2)(r^2+a_3^2)-q(a_1a_2a_3g-r^2)}\,,\nn\\
w_2=\frac{a_2(r^2+a_1^2)(r^2+a_3^2)(1+g^2r^2)-g q(a_1a_3-a_2
g r^2)}{(r^2+a_1^2)(r^2+a_2^2)(r^2+a_3^2)-q(a_1a_2a_3g-r^2)}\,,\nn\\
w_3=\frac{a_3(r^2+a_1^2)(r^2+a_2^2)(1+g^2r^2)-g q(a_1a_2-a_3 g
r^2)}{(r^2+a_1^2)(r^2+a_2^2)(r^2+a_3^2)-q(a_1a_2a_3g-r^2)}\,.\eea
Note $\Omega_1=w_1(r_0)$, $\Omega_2=w_2(r_0)$ and
$\Omega_3=w_3(r_0)$ are the three angular velocities of the black
hole. It is obvious that the result agrees with
(\ref{key.result}). In the limit $r\rightarrow+\infty$, we find
\be w_1=g^2a_1+\cO(\frac1r)\,,\quad
w_2=g^2a_2+\cO(\frac1r)\,,\quad w_3=g^2a_3+\cO(\frac1r)\,.\ee
In the case $g=0$, the solution is asymptotically flat, and we
have $\xi=\pd_t$ as $r\rightarrow+\infty$.

\subsection{d=8}

In arbitrary dimensions, there is the Kerr-NUT-AdS solution found
in \cite{chen.lu.pope06b}. Still we are not able to discuss the
general case of arbitrary dimensions. Here we will only consider
the case of eight dimensions (lower dimension ones are already
covered in the previous examples). Since the metrics obtained in
\cite{chen.lu.pope06b} are already of the form
(\ref{metric.general}), we do not need to repeat the calculation
done in the last section. By comparing equation (13) in
\cite{chen.lu.pope06b} with our metric (\ref{metric.8d}), we find
that
\be w_\alpha=\frac{a_\alpha(1+g^2r^2)}{r^2+a_\alpha^2}\,,\quad
\alpha=1,2,3\,.\ee
In the limit $r\rightarrow+\infty$, we find
\be w_\alpha=g^2a_\alpha+\cO(\frac1r)\,,\quad \alpha=1,2,3\,.\ee
In the case $g=0$, the solution is asymptotically flat, and we
have $\xi=\pd_t$ as $r\rightarrow+\infty$.

It is straightforward to generalize the calculation to higher
dimensions, and we expect that the basic properties of the result
stay the same.

\subsection{Summary of the section}

To summarize this section, we have shown that, when the black hole
solutions are asymptotically flat and when the coordinate system
is asymptotically static, then the vector field (\ref{key.result})
approaches the time Killing vector at the spatial infinity.
Because of technical reasons, we have not been able to do the
calculation for several interesting examples. But it is quite
likely that result \#2 will hold for all stationary black holes
that have a well defined vielbein expression outside the black
hole horizon.

\section{Summary}

In this paper, we have constructed a vector field by using the
``conserved current" of a particular spinor field. We have shown
that, in the background of a stationary black hole, the vector
field always approaches the null Killing vector on horizon. When
the black hole is asymptotically flat and when the coordinate
system is asymptotically static, the same vector field also
becomes the time Killing vector at the spatial infinity. The
required constraint on the spinor field is simple and universal
(valid for any spacetime dimensions).

It is still not clear as to the physical nature of the vector
field or the corresponding spinor field. Our original motivation
for studying the vector field was to construct a possible fluid
flow underlying the spacetime \cite{mei11}. For asymptotically
flat black hole solutions, the behavior of the vector field fits
very well with our intuitive picture about the speculated fluid
that may underly our spacetime. One can imagine that the fluid is
dragged by the black hole horizon (Hence the same velocity on the
horizon\footnote{Here we simply let $U=\xi$, where $U^\mu$ is the
velocity of the fluid. We no longer assume $U^2=-1$ as was done in
\cite{mei11}.}). Then the angular velocity steadily decreases
until it vanishes at the spatial infinity. However, such a picture
is still highly hypothetical, and one should be open minded
towards other possible explanations.

Another interesting possibility is to treat (\ref{key.result}) as
sort of generalized angular velocity function. This may be offer
some insight towards peculiar objects such as the BTZ black hole
\cite{btz92,carlip05}. In particular, one may say that the black
hole now has a finite angular velocity at the spatial infinity,
even though the horizon angular velocity is zero. Again, much more
work is needed before one can take such a possibility seriously.

Regardless of what the physical interpretation may be, it is
unexpected and also quite amazing that something like
(\ref{key.constr}) and (\ref{key.result}) can exist. Given the
remarkable features summarized in {\bf Result \#1} and {\bf Result
\#2}, it will be very interesting to see possible applications of
the vector field (\ref{key.result}), or the corresponding spinor
field (\ref{key.constr}), or both.

\section*{Acknowledgement}

This work was supported by the Alexander von Humboldt-Foundation.


\end{document}